\documentclass[12pt]{article}
\usepackage{amsthm,amsfonts,amsmath,amssymb,mathrsfs,graphicx,url}
\usepackage[usenames,dvipsnames]{color}

\title{Multi-Time Wave Functions versus\\ Multiple Timelike Dimensions}

\author{
Matthias Lienert,\footnote{Department of Mathematics,
     Rutgers University,
     110 Frelinghuysen Road, Piscataway, NJ 08854-8019, USA.
     E-mail: mlienert@math.rutgers.edu}\ \ 
S\"oren Petrat\footnote{Department of Physics, Jadwin Hall, Princeton University, 
	Washington Road, Princeton, NJ 08544-0708, USA. E-mail: spetrat@princeton.edu},\ \ and 
Roderich Tumulka\footnote{Fachbereich Mathematik, Eberhard-Karls-Universit\"at, Auf der Morgenstelle 10, 72076 T\"ubingen, Germany. E-mail: roderich.tumulka@uni-tuebingen.de}
}
\date{September 19, 2017}

\newcommand{\be}{\begin{equation}}
\newcommand{\ee}{\end{equation}}
\newcommand{\Hilbert}{\mathscr{H}}

\newcommand{\RRR}{\mathbb{R}}

\newcommand{\CCC}{\mathbb{C}}

\newcommand{\scp}[2]{\langle #1|#2 \rangle}

\newcommand{\vx}{\boldsymbol{x}}

\newcommand{\sS}{\mathscr{S}}

\begin{document}

\maketitle

\begin{abstract}
Multi-time wave functions are wave functions for multi-particle quantum systems that involve several time variables (one per particle). In this paper we contrast them with solutions of wave equations on a space-time with multiple timelike dimensions, i.e., on a pseudo-Riemannian manifold whose metric has signature such as ${+}{+}{-}{-}$ or ${+}{+}{-}{-}{-}{-}{-}{-}$, instead of ${+}{-}{-}{-}$. Despite the superficial similarity, the two behave very differently: Whereas wave equations in multiple timelike dimensions are typically mathematically ill-posed and presumably unphysical, relevant Schr\"odinger equations for multi-time wave functions possess for every initial datum a unique solution on the spacelike configurations and form a natural covariant representation of quantum states. 

\medskip

\end{abstract}

\section{Introduction}

In reaction to recent developments about multi-time wave functions \cite{pt:2013a,pt:2013c,pt:2013e,pt:2013d,lienert:2015a,lienert:2015b,lienert:2015c,LN:2015,ND:2016}, some colleagues have expressed concern that this approach might be fundamentally misguided because, on the one hand, multi-time wave functions such as
\be\label{phi}
\phi(t_1,\vx_1,\ldots,t_n,\vx_n)
\ee
with $\vx_j\in \RRR^3$ and $t_j\in \RRR$ for $n$ particles are functions on a space $\RRR^{4n}$ with several time variables, and, on the other hand, there are severe difficulties with the superficially similar situation of the wave equation or Dirac equation on pseudo-Riemannian manifolds with several timelike dimensions. For example, in a review of the article \cite{pt:2013e} for \textit{Mathematical Reviews}, Bera \cite{bera} warned his readers about multi-time wave functions:
\begin{quotation}
Overall I want to say that the wave equation with more than one time dimension is ill-posed in the sense of Hadamard. The recent work of Weinstein and Craig has presented a sufficient constraint on such problems endowing them with well-posedness conditions near those of standard single time dimension problems. Physical theories or interpretations with multiple time dimensions are usually shunned by most researchers. The standard consensus among physicists is that such problems are unstable. This
interpretation has also spread to the mathematical community, and so such problems are generally not considered.
\end{quotation}
That is, Bera is saying, multi-time wave functions should be expected to be ill-behaved, and that is why physicists as well as mathematicians are not considering these objects.\footnote{An initial-value problem for a PDE is called \emph{well-posed} (in the sense of Hadamard) \cite{Evans:2010} if for every initial datum (from an appropriate function space) a solution exists for all times, is unique, and depends continuously on the initial datum.}

The purpose of this note is to explain why for multi-time wave functions there is, actually, no reason for concern: they are not at all ill-behaved in the way Bera worried about. In fact, the situation of multi-time wave functions is very different, perhaps surprisingly different, from that of the wave equation in a space-time with multiple timelike dimensions. It is the latter, not the former, that is ill-posed. In contrast, the relevant equations for multi-time wave functions do possess a unique solution for every initial condition, so that the multi-time evolution is well defined, no instability occurs, and the difficulties pointed out by Craig and Weinstein \cite{CW:2009} do not apply. 
A general overview of multi-time wave functions can be found in \cite{LPT:2017}.

To give names to the two things we are comparing, let us call the first thing, the evolution equation for a multi-time wave function, a \emph{multi-time Schr\"odinger equation} or MTS equation, and the second thing an equation \emph{with multiple timelike dimensions}, or MTD equation. Our first message in this letter is that 
\[
\text{while MTD is ill-posed, MTS is well-posed.}
\]
Another message goes beyond the mathematical properties and concerns the physical meaning: While MTD is rather speculative and not based on empirical evidence, MTS is a natural covariant version of the non-relativistic wave function and directly connected to probabilities of experimental outcomes. 

Let us write down explicit examples of such equations.

\section{Different Equations, Different Solutions}

We set $\hbar = 1 = c$. Let $\tilde g^{\mu\nu}$ be a metric with $p\geq 2$ spacelike and $q\geq 2$ timelike dimensions,
\be\label{gmunu}
\tilde g^{\mu\nu} = \mathrm{diag}(\underbrace{+1,\ldots,+1}_{q},\underbrace{-1,\ldots,-1}_{p})
\ee
on $\RRR^{p+q}$. The second-order ``$p+q$'' Klein--Gordon equation reads
\be\label{KG}
\tilde g^{\mu\nu}\partial_\mu\partial_\nu \psi = m^2 \psi
\ee
with $\psi: \RRR^{p+q}\to \CCC$; it is called an ``ultrahyperbolic equation.'' The case $m=0$ corresponds to the ``$p+q$'' wave equation.
Analogously, one can consider a ``$p+q$'' Dirac equation associated with $\tilde g^{\mu\nu}$,
\be\label{Dirac}
i\tilde\gamma^\mu \partial_\mu \psi = m\psi\,,
\ee
where $\psi: \RRR^{p+q} \to \CCC^k$, and the $\tilde\gamma^\mu$ are $k\times k$ matrices that realize the Clifford algebra relations
\be\label{Clifford}
\tilde\gamma^\mu\tilde\gamma^\nu + \tilde\gamma^\nu \tilde\gamma^\mu = 2\tilde g^{\mu\nu} \, I\,.
\ee
So, \eqref{KG} and \eqref{Dirac} are examples of MTD equations. Multiple timelike dimensions have actually been proposed to be realized in nature in the context of string theory and extra-dimensional high energy physics (see, e.g., \cite{Tegmark:1997,Sparling:2007,PRS:2016} and references therein).

\bigskip

MTD equations should be contrasted with the system of Schr\"odinger equations for a multi-time wave function $\phi$ as in \eqref{phi}, which is of the form \cite{dirac:1932,dfp:1932,bloch:1934}
\be\label{dtj}
i \partial_{t_j} \phi = H_j\, \phi \qquad \forall j=1,\ldots,n
\ee
with the ``partial Hamiltonian'' $H_j$. The simplest examples of such MTS equations \cite{schweber:1961} correspond to non-interacting particles, for example with $H_j$ the Hamiltonian of the $3+1$ Dirac equation acting on particle $j$,
\be\label{multiDirac}
H_j= -i \sum_{a=1}^3 \alpha_{aj} \partial_{x^a_{j}} + \beta_j m  \,,
\ee
where $\alpha_{aj}=\gamma^0\gamma^a$ and $\beta_j = \gamma^0$ act on the spin index of particle $j$. The MTS equations arise from the desire to express entangled multi-particle quantum states through a wave function in a covariant way, as discussed in Section~\ref{sec:significance} below. (It is not assumed that space-time has any other signature than ${+}{-}{-}{-}$.)

\bigskip

The most basic difference between MTS and MTD is that the MTS equation \eqref{dtj} is a system of $n$ equations per component of $\phi$, whereas the MTD equation \eqref{KG} [or \eqref{Dirac}] is just one equation per component of $\psi$.

Here is another difference. The relevant comparison concerns the case of a $4n$-dimensional metric $\tilde g_{\mu\nu}$ with $q=n$ timelike and $p=3n$ spacelike dimensions. Then in the MTD equation \eqref{KG} [or \eqref{Dirac}], all $4n$ derivatives appear, in fact in a rather symmetric way. In the MTS equation \eqref{dtj} for index $j$, in contrast, only 4 derivatives appear, viz., those relative to $x_j$.

Given that MTS and MTD are different equations, it is not surprising that they behave differently concerning the existence and uniqueness of solutions. Or perhaps it is \emph{still} surprising, in view of the fact that they both involve several time variables. So let us have a closer look at why MTS is well-posed and MTD ill-posed.

\section{Behavior of Each Equation}

Already a mere count of dimensions and equations suggests that MTS may be well-posed but MTD cannot be for initial data at time 0. The MTS equation \eqref{dtj} provides $n$ equations for each component of $\phi$, one for each time variable; this suggests that initial data for $\phi$, specified on the $3n$-dimensional surface where all time variables are set to 0, may determine $\phi$ for all values of the time variables. In contrast, the one MTD equation \eqref{KG} or \eqref{Dirac} could not be expected to determine more than the dependence on one variable, say $t_1$, so the initial data, in order to determine the solution $\psi$ uniquely, would have to be specified on the $(4n-1)$-dimensional surface where $t_1=0$ and would in particular specify the depedence of $\psi$ on $t_2,\ldots,t_n$ for $t_1=0$. This means that initial data at time $0$ (i.e., all time variables equal to 0) are insufficient to determine the solution, so uniqueness of the solution fails. The rigorous analysis of Craig and Weinstein \cite{CW:2009} confirms this picture for the MTD equation \eqref{KG}, except that the situation is even a little worse: even if we specify initial data on the $(4n-1)$-dimensional surface where $t_1=0$, then the continuous dependence of the solution on the initial datum (in a norm analogous to the one commonly used in the case with only one timelike dimension) holds only for special initial data, viz., those whose Fourier transform (in $4n-1$ variables) vanishes on wave vectors $k$ that are timelike relative to $\tilde{g}^{\mu\nu}$. This is the kind of unstable behavior that Bera referred to in \cite{bera}. 

Returning to the MTS equation for non-interacting particles according to \eqref{multiDirac}, we can be explicit about what the solution actually looks like. In this case, $H_j$ can be regarded as a self-adjoint operator on the space $\Hilbert=L^2(\RRR^{3n},(\CCC^4)^{\otimes n})$, and $\phi$, as a function defined on the space $\RRR^n=\{(t_1,\ldots,t_n)\}$ spanned by the time axes and with values in $\Hilbert$, is given by \cite{schweber:1961}
\be
\phi(t_1,\ldots,t_n) = e^{-iH_1 t_1} \cdots e^{-iH_n t_n} \phi(0,\ldots,0)\,.
\ee
Thus, the solution exists, is unique, and depends continuously on the initial datum $\phi(0,\ldots,0)$ in the same $L^2$ sense as for any Schr\"odinger equation.

The situation becomes more involved in the presence of interaction. Since we are trying to simultaneously solve $n$ PDEs \eqref{dtj} governing different derivatives, we can only expect solutions to exist for arbitrary initial conditions (or from a reasonably large class) if the compatibility or consistency condition
\be\label{consistency}
\Bigl[ i\partial_{t_j} - H_j, i \partial_{t_k} - H_k \Bigr] =0
\ee
is satisfied. This has long been known \cite{bloch:1934}; see \cite{pt:2013a} for a detailed discussion. The consistency condition is automatically satisfied for \emph{non}-interacting particles, but it is less obvious how to set up MTS equations \emph{with interaction} that satisfy it. Nevertheless, it can be done \cite{bloch:1934,DV81,DV82b,CVA:1983,DV85,ML91,pt:2013a,pt:2013c,pt:2013d,lienert:2015a,lienert:2015b,lienert:2015c,LN:2015,ND:2016}, with $\phi$ defined on the spacelike configurations, and the authors of these references have all been completely aware of the consistency condition \eqref{consistency}. In fact, they have used it as a guideline for how to set up MTS equations. 

The upshot is that the MTS equations considered in the references just cited are well-posed; this has been shown rigorously in some cases \cite{pt:2013a,lienert:2015a,lienert:2015b,lienert:2015c,LN:2015}, and made plausible in others \cite{DV81,DV82b,DV85,pt:2013c,pt:2013d,ND:2016}. In some of the latter cases, the obstacle to a rigorous proof lies in the ultraviolet divergence, which affects also the single-time theory and is unrelated to the multi-time approach.

Readers may still wonder how well-posedness of  MTS can fit together with ill-posedness of MTD for the following reason. For $n$ non-interacting Klein-Gordon particles with equal mass, the partial Hamiltonians
\be
H_j =\sqrt{-\Delta_j +m^2}
\ee
with $\Delta_j = \sum_{a=1}^3 \partial_{x_j^a}^2$ commute pairwise and are time independent, so the consistency condition \eqref{consistency} is fulfilled, so a unique solution $\phi$ of the MTS equation \eqref{dtj} exists for every initial datum. As a consequence of \eqref{dtj}, this $\phi$ satisfies
\be\label{KGj}
\Bigl( \partial_{x_j^0}^2 - \sum_{a=1}^3 \partial_{x_j^a}^2 \Bigr) \phi = m^2 \phi
\ee
and thus, by summing \eqref{KGj} over $j$, satisfies also the MTD equation \eqref{KG} with $q=n$, $p=3n$, and $m$ rescaled to $\sqrt{n} \, m$. So how can MTD equations be bad?

The first and foremost aspect is that the main trouble with MTD equations was the lack of uniqueness, and the solutions $\phi$ just constructed are simply special solutions, obtained from a more restrictive set of equations \eqref{dtj}. A further aspect concerns the continuous dependence on initial data: the norm to which continuity refers would naturally be chosen differently depending on whether we regard the $(4n-1)$-dimensional set $\{t_1=0\}$ or the $3n$-dimensional set $\{t_1=t_2=\ldots = t_n=0\}$ as the set where initial data are specified. Simultaneous solutions of \eqref{KGj} for all $j$ have infinite norm in the norm used by Craig and Weinstein \cite{CW:2009}.

\section{Physical Significance}
\label{sec:significance}

The hypothesis of MTD may be intriguing, but it is rather speculative and not empirically grounded. In contrast, the multi-time wave functions $\phi(t_1,\vx_1,\ldots,t_n,\vx_n)$ of MTS equations are the relativistic analog of ordinary non-relativistic multi-particle wave functions $\psi(\vx_1,\ldots, \vx_n,t)$. 
There are several reasons supporting this view.

\begin{itemize}
\item 
When we consider a configuration such as $(\vx_1,\ldots,\vx_n)$ at time $t$ then we really mean $n$ space-time points $(t,\vx_1),\ldots,(t,\vx_n)$. In another Lorentz frame differing from the first by a Lorentz transformation $\Lambda$, these points have coordinates
\be
(t_1',\vx_1')=\Lambda(t,\vx_1),\ldots,(t_n',\vx_n')=\Lambda(t,\vx_n)
\ee
which are in general no longer simultaneous, $t_i'\neq t_j'$. One is thus naturally led to considering a wave function $\phi$ on the set
\be
\sS= \Bigl\{ (x_1,\ldots, x_n)\in (\RRR^4)^n: (x_i-x_j)^\mu(x_i-x_j)_\mu <0 \: \forall i\neq j \Bigr\}
\ee
of all spacelike configurations (where the Minkowski metric has signature ${+}{-}{-}{-}$). Needless to say, $\phi$ has several time variables.

\item It turns out (e.g., \cite{LPT:2017,pt:2013a}) that the single-time wave function $\psi$ of any fixed Lorentz frame and $\phi$ are related according to
\be
\psi(\vx_1,\ldots,\vx_n,t) = \phi(t,\vx_1,\ldots, t,\vx_n)\,.
\ee
That is, for simultaneous configurations, $\phi$ agrees with the usual kind of wave function $\psi$.

\item 
 $\phi$ is directly related to detection probabilities as follows \cite{bloch:1934,LT:2017}. Consider for simplicity $n=2$. For two non-interacting particles,  if we measure the position of particle 1 at time $t_1$ (in a certain Lorentz frame) and that of particle 2 at time $t_2$, then the joint probability density of the two outcomes is $|\phi(t_1,\vx_1,t_2,\vx_2)|^2$ \cite{bloch:1934}. Also, if we place detectors along a spacelike hypersurface $\Sigma$, then the joint distribution of the points $x_1,x_2\in\Sigma$ where the two particles get detected has density $\rho(x_1,x_2)$ (relative to the notion of volume defined by the 3-metric on $\Sigma$) given by the appropriate version of $|\cdot|^2$ applied to $\phi(x_1,x_2)$ \cite{LT:2017}, viz.,
\be
\rho(x_1,x_2) = \overline{\phi}(x_1,x_2) \: [\nu(x_1) \otimes \nu(x_2) ] \: \phi(x_1,x_2)\,,
\ee
where $\nu(x)$ is the matrix in spin space representing the scalar product in the Lorentz frame tangent to $\Sigma$ in $x$; for example, for Dirac wave functions, $\nu(x) = n_\mu(x) \, \gamma^\mu$ with $n_\mu(x)$ the future unit normal vector to $\Sigma$ at $x$.

\item For relevant models, the time evolution of $\phi$ is governed by rather simple PDEs, the MTS equations, which play a role analogous to the Schr\"odinger equation for $\psi$. Examples of such PDEs are described in \cite{LPT:2017} and discussed in detail in \cite{pt:2013c,lienert:2015a,LN:2015}; they are of the form \eqref{dtj}, with one equation for each time variable. The solution $\phi$ of these PDEs is a Lorentz covariant object, one that can be expressed in any Lorentz frame.

\item Also for quantum field theories, one can express the quantum state as a wave function, even if many texts leave this picture aside. In a non-relativistic setting, elements $\psi$ of Fock space can be regarded as functions on the configuration space $\Gamma_o(\RRR^3)$ of a variable number of particles, 
\be
\Gamma_o(\RRR^3) = \bigcup_{n=0}^\infty (\RRR^3)^n\,,
\ee
so $\psi = (\psi^{(0)},\psi^{(1)}, \psi^{(2)},\ldots)$ with the $n$-particle sector $\psi^{(n)}$ an $n$-particle wave function on $\RRR^{3n}$, normalized so that
\be
\|\psi\|^2 := \sum_{n=0}^\infty \|\psi^{(n)}\|^2 =1\,. 
\ee
Then, $\psi$ can be expressed in terms of the field operators $\Phi(\vx)$, $\vx\in \RRR^3$, according to
\be
\psi^{(n)}(\vx_1,\ldots,\vx_n) = \scp{\emptyset}{\Phi(\vx_1)\cdots \Phi(\vx_n)|\Psi}
\ee
for mutually distinct $\vx_j\in \RRR^3$, $|\emptyset\rangle$ the Fock vacuum, and $|\Psi\rangle$ a vector in Fock space.

Now it turns out \cite{pt:2013c} that the multi-time wave function $\phi$ satisfies an analogous relation to the field operators, viz.,
\be\label{phiPhi}
\phi^{(n)}(x_1,\ldots, x_n) = \scp{\emptyset}{\Phi(x_1) \cdots \Phi(x_n)| \Psi}
\ee
for mutually spacelike and distinct $x_j \in \RRR^4$ and with $\Phi(x_j)$ the time evolved field operator in the Heisenberg picture. This is another reason for regarding $\phi$ as a natural extension of $\psi$. The canonical (anti-)commutation relations for $\Phi(x)$ imply that $\phi^{(n)}$ is (anti-)symmetric against permutation of the space-time points $x_j$---the natural extension of the bosonic (respectively, fermionic) symmetry of $\psi$ against permutation of the space points $\vx_j$.
\end{itemize}

\section{Outlook}

Relativistic quantum physics usually means quantum field theory (QFT), and in QFT it is not so common to talk about wave functions. Still, one may wish to express the quantum state as a wave function also in QFT (an approach used, e.g., by Schweber \cite{schweber:1961}), in order to emphasize the similarity with quantum mechanics and to make visible the relation between the state vector and the space-time points (whereas in the Heisenberg picture, the state vector is a more abstract thing). Further benefits include a direct expression for detection probabilities in terms of $|\psi|^2$ (or $|\phi|^2$) as well as the possibility of expressing boundary conditions through simple equations on $\psi$ (or $\phi$), including the interior-boundary conditions \cite{ML91,LSTT} used for defining particle creation at a point-shaped source.

So, an equivalent reformulation of QFT in terms of wave functions is possible when the Hilbert space is a Fock space or a tensor product of Fock spaces, and the relativistic version of that involves a multi-time wave function $\phi$. This approach can also be of interest as another way of deriving QFT starting from quantum mechanics, by adding particle creation and annihilation and introducing relativistic Hamiltonians. This could be regarded as a pedestrian derivation of QFT, in which state vectors are not just elements of an abstract Hilbert space but concretely realized as wave functions depending on space-time points, while field operators are not introduced by axiom but are obtained from creation and annihilation operators acting on wave functions. As a consequence, in this approach one would indeed  take \eqref{phiPhi} as a theorem and not as the definition of $\phi$. Instead, $\phi$ is defined as the solution of the MTS equations \eqref{dtj}, so that the existence and uniqueness of solutions becomes a central issue. As we have argued, existence and uniqueness do hold in exactly the way one would hope for, much in contrast to the situation of MTD equations \eqref{KG} and \eqref{Dirac}.

\bigskip

\noindent{\it Acknowledgments.} We thank an anonymous referee for her or his comments on a previous version of this note.\\[2mm]
\begin{minipage}{15mm}
\includegraphics[width=13mm]{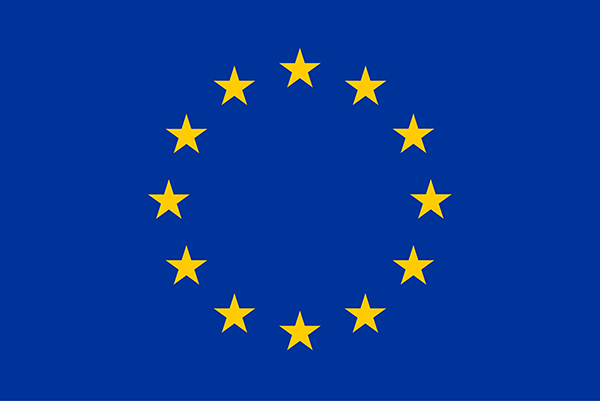}
\end{minipage}
\begin{minipage}{143mm}
This project has received funding from the European Union's Framework for Research and Innovation Horizon 2020 (2014--2020) under the Marie Sk{\l}odowska-
\end{minipage}\\[1mm]
Curie Grant Agreement No.~705295. 
S.P.\ gratefully acknowledges support from the German Academic Exchange Service (DAAD).


\begin{thebibliography}{19}

\bibitem{bera}
P.~K.~Bera:
\newblock Review of \cite{pt:2013e}.
\newblock {\it Mathematical Reviews} 3165768 (2015)

\bibitem{bloch:1934}
F.~Bloch:
\newblock {Die physikalische Bedeutung mehrerer Zeiten in der
  Quantenelektrodynamik}.
\newblock {\em Physikalische Zeitschrift der Sowjetunion} {\bf 5}: 301--305 (1934)

\bibitem{CW:2009}
W.~Craig and S.~Weinstein:
\newblock On determinism and well-posedness in multiple time dimensions.
\newblock {\it Proceedings of the Royal Society A} {\bf 465}: 3023--3046 (2009)
\newblock \url{http://arxiv.org/abs/0812.0210}

\bibitem{CVA:1983}
H.W.~Crater and P.~Van Alstine:
\newblock Two-body Dirac equations.
\newblock {\it Annals of Physics} {\bf 148}: 57--94 (1983)

\bibitem{dirac:1932}
P.A.M. Dirac:
\newblock {Relativistic Quantum Mechanics}.
\newblock {\em Proceedings of the Royal Society London A} {\bf 136}: 453--464 (1932)

\bibitem{dfp:1932}
P.A.M. Dirac, V.A. Fock, and B.~Podolsky:
\newblock {On Quantum Electrodynamics}.
\newblock {\em Physikalische Zeitschrift der Sowjetunion} {\bf 2(6)}: 468--479 (1932).
\newblock Reprinted in J. Schwinger: {\em Selected Papers on Quantum
  Electrodynamics}, New York: Dover (1958)

\bibitem{DV81}
Ph.~Droz-Vincent:
\newblock Relativistic Wave Equations for a System of Two Particles with Spin $\tfrac{1}{2}$.
\newblock \textit{Lettere al Nuovo Cimento} \textbf{30}: 375--378 (1981)

\bibitem{DV82b} 
Ph.~Droz-Vincent:  
\newblock Second quantization of directly interacting particles. 
\newblock Pages 81--101 in J.~Llosa (ed.): \textit{Relativistic Action at a Distance: Classical and Quantum Aspects}, Berlin: Springer-Verlag (1982)

\bibitem{DV85} 
Ph.~Droz-Vincent: 
\newblock Relativistic quantum mechanics with non conserved number of particles. 
\newblock \textit{Journal of Geometry and Physics} {\bf 2(1)}: 101--119 (1985)

\bibitem{Evans:2010}
L.C.~Evans:
\newblock {\em {Partial Differential Equations}}.
\newblock American Mathematical Society, Providence, Rhode Island, second edition (2010)

\bibitem{LSTT} 
J.~Lampart, J.~Schmidt, S.~Teufel, and R.~Tumulka: 
\newblock Particle Creation at a Point Source by Means of Interior-Boundary Conditions.
\newblock Preprint (2017)
\newblock \url{https://arxiv.org/abs/1703.04476}

\bibitem{lienert:2015a}
M.~Lienert:
\newblock A relativistically interacting exactly solvable multi-time model for two mass-less Dirac particles in 1+1 dimensions.
\newblock {\it Journal of Mathematical Physics} {\bf 56}: 042301 (2015)
\newblock \url{http://arxiv.org/abs/1411.2833}

\bibitem{lienert:2015b}
M.~Lienert:
\newblock On the question of current conservation for the Two-Body Dirac equations of constraint theory.
\newblock {\it Journal of Physics A: Mathematical and Theoretical} {\bf 48}: 325302 (2015)
\newblock \url{http://arxiv.org/abs/1501.07027}

\bibitem{lienert:2015c}
M.~Lienert:
\newblock Lorentz invariant quantum dynamics in the multi-time formalism.
\newblock Ph.D.~thesis, Mathematics Institute, Ludwig-Maximilians University, Munich, Germany (2015)

\bibitem{LN:2015}
M.~Lienert and L.~Nickel:
\newblock A simple explicitly solvable interacting relativistic $N$-particle model.
\newblock {\it Journal of Physics A: Mathematical and Theoretical} {\bf 48}: 325301 (2015) 
\newblock \url{http://arxiv.org/abs/1502.00917}

\bibitem{LPT:2017}
M.~Lienert, S.~Petrat, and R.~Tumulka:
\newblock Multi-Time Wave Functions.
\newblock {\it Journal of Physics: Conference Series} {\bf 880}: 012006 (2017)
\newblock \url{http://arxiv.org/abs/1702.05282}

\bibitem{LT:2017}
M.~Lienert and R.~Tumulka:
\newblock Born's Rule on Arbitrary Cauchy Surfaces.
\newblock \url{http://arxiv.org/abs/1706.07074}

\bibitem{ML91} M.~Moshinsky and G.~L\'opez Laurrabaquio:
\newblock  Relativistic interactions by means of boundary conditions: The Breit--Wigner formula.
\newblock  \textit{Journal of Mathematical Physics} \textbf{ 32}: 3519--3528 (1991)

\bibitem{ND:2016}
L.~Nickel and D.-A.~Deckert:
\newblock Consistency of multi-time Dirac equations with general interaction potentials.
\newblock \textit{Journal of Mathematical Physics} {\bf 57}: 072301 (2016)
\newblock \url{http://arxiv.org/abs/1603.02538}

\bibitem{pt:2013a}
S.~Petrat and R.~Tumulka:
\newblock Multi-Time Schr\"odinger Equations Cannot Contain Interaction Potentials.
\newblock \textit{Journal of Mathematical Physics} {\bf 55}: 032302 (2014)
\newblock \url{http://arxiv.org/abs/1308.1065}

\bibitem{pt:2013c}
S.~Petrat and R.~Tumulka:
\newblock Multi-Time Wave Functions for Quantum Field Theory.
\newblock {\it Annals of Physics} {\bf 345}: 17--54 (2014)
\newblock \url{http://arxiv.org/abs/1309.0802}

\bibitem{pt:2013e}
S.~Petrat and R.~Tumulka:
\newblock Multi-Time Equations, Classical and Quantum.
\newblock \textit{Proceedings of the Royal Society A} {\bf 470(2164)}: 20130632 (2014)
\newblock \url{http://arxiv.org/abs/1309.1103}

\bibitem{pt:2013d}
S.~Petrat and R.~Tumulka:
\newblock Multi-Time Formulation of Pair Creation.
\newblock \textit{Journal of Physics A: Mathematical and Theoretical} {\bf 47}: 112001 (2014)
\newblock \url{http://arxiv.org/abs/1401.6093}

\bibitem{PRS:2016}
E.~Piceno, A.~Rosado, and E.~Sadurn\'\i:
\newblock Fundamental constraints on two-time physics.
\newblock {\it European Physical Journal Plus} {\bf 131}: 352 (2016)
\newblock \url{http://arxiv.org/abs/1512.05345}

\bibitem{schweber:1961}
S.~Schweber:
\newblock {\em {An Introduction To Relativistic Quantum Field Theory}}.
\newblock Row, Peterson and Company (1961)

\bibitem{Sparling:2007}
G.A.J.~Sparling:
\newblock Germ of a synthesis: space--time is spinorial, extra dimensions are time-like.
\newblock {\it Proceedings of the Royal Society A} {\bf 463}: 1665--1679 (2007)
% (doi:10.1098/rspa.2007.1839)

\bibitem{Tegmark:1997}
M.~Tegmark:
\newblock On the dimensionality of space-time. 
\newblock {\it Classical and Quantum Gravity} {\bf 14}: L69--L75 (1997)
%(doi:10.1088/0264-9381/14/4/002)
\newblock \url{http://arxiv.org/abs/gr-qc/9702052}

\end{thebibliography}
\end{document}